\documentclass{ws-p8-50x6-00}

\begin{document}

\title{Problems and cures (partial) for holographic cosmology}

\author{M.A. Per and A. Segui}

\address{Departamento de F\'{\i}sica Te\'orica. Facultad de Ciencias.
Universidad de Zaragoza. 50009-Zaragoza, Spain\\ 
E-mail: segui@unizar.es}




\maketitle

\abstracts{
We analyze the validity of the generalized covariant entropy bound near
the apparent horizon of isotropic expanding cosmological models.
We encounter violations of the bound for cosmic times smaller
than a threshold. By introducing an infrared cutoff we are able to mantain
the bound for a radiation dominated universe.
We study different physical mechanisms to restore the
bound, as a non-additivity of the entropy at a fundamental level and/or
a cosmological uncertainty relation.}

\section{Introduction}
The very nature of the theory of quantum gravity is being revealed
by consistent formulations of quantum gravitational phenomena
as the string/M theory do, as well as by
{\em principles} that underlie the precise formulation of the theory.
It is believed that a primary principle of the theory of quantum gravity
is the Holographic Principle, that states that a
physical system can be described only
by degrees of freedom living in its boundary. When we
descend to a classical description the previous principle adopts the
form of an entropy bound that admits different formulations depending
on the strength of the gravitational interaction. 

There was a proliferation of entropy
bounds that made any prediction meaningless. Fortunately a covariant
formulation for the maximum entropy allowed for a physical system was
given\cite{bousso} and the possibilities drastically reduced. 
In order that different
observers agree with the entropy of a system they must measure the entropy
traversing a {\em light sheet} (LS), that is the locus spanned by the
null congruence, generated with the 
area decreasing light rays orthogonal to a spacelike
codimension two surface. The LS ends in a singularity, or a caustic
where the area begins to increase. The covariant 
entropy bound (CEB) establishes that
the amount of entropy measured in this way is bounded by one quarter of 
the area, in Planck units, of
the spacelike codimension two surface where the congruence begins. 
All the entropy bounds were special relaxed cases of the
covariant bound, except the Bekenstein bound\cite{bekenstein} (BB). 

To put
all the entropy bounds together we need to use the generalized covariant
entropy bound\cite{wald}(GCEB) that truncates the LS with a second
spacelike codimension two surface and establish that the amount of entropy
traversing the truncated LS is bounded by one quarter of 
the difference of
area between both boundaries. It is evident that the GCEB implies the CEB, and
it has been shown\cite{bousso2} that the BB can 
also be deduced from the GCEB. So the
GCEB is the stronger formulation of the bound and by imposing it on 
different physical systems we obtain further insights on the nature of 
the holographic principle and the theory of quantum gravity.

We are interested on the validity of the GCEB for expanding isotropic
cosmological scenarios. It is the case that the GCEB is violated near the 
apparent horizon (AH)\cite{guedens}{\footnotesize'}\cite{bousso3}. 
The reason is that on the AH the LS 
develops a maximum area; if we truncate the LS near the AH the difference of
areas goes to zero as the second power of the affine parameter that 
parametrizes the LS, whereas the entropy traversing the LS goes to zero 
as the first power of the same parameter; so unavoidably we are facing 
a violation of the GCEB for small enough values of the affine parameter.
In\cite{bousso3} a possible resolution of the violation of the GCEB is 
addressed by admitting for the carriers of entropy, only particles with 
a wavelength smaller than the physical separation between the two spacelike
surfaces bounding the LS. 

In this article we study the difficulties that appear near the AH to 
satisfy the GCEB in Friedman-Lema\^{\i}tre-Robertson-Walker(FLRW) cosmologies.
We can solve the discrepancies using an infrared cutoff to cut the modes
with a wavelength larger than the size they traverse; we do it for a 
radiation dominated model. The use of different cutoffs for different parts
of a given LS seems to be related with an intrinsic non additivity of the 
entropy that we comment. Because we use the cosmic time 
as the affine parameter, we are able to obtain an expression 
for the amount of time we must wait since the
AH to satisfy the GCEB. Our result can be put in a suggestive form as a 
cosmological uncertainty relation.

.

\section{The GCEB in a FLRW universe}
\subsection{The location of the AH in proper coordinates}
\begin{figure}[t]
\epsfxsize=25pc 
\epsfbox{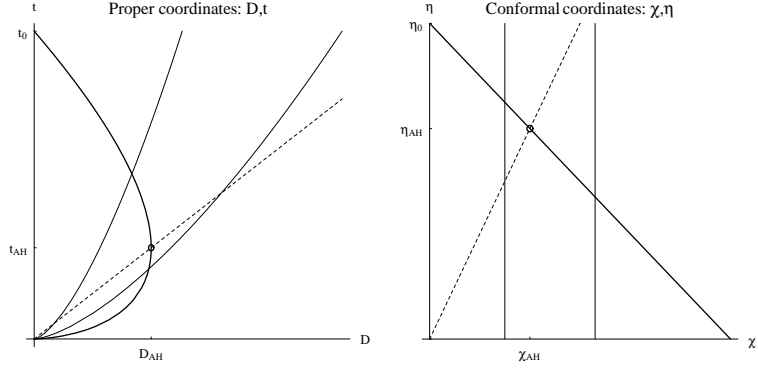} 
\caption{An ingoing null geodesic is depicted for the two coordinates,
proper and conformal. The dashed line shows the position of the AH.
The timelike geodesics of two galaxies are also shown.   \label{coord}}
\end{figure}

Now we study the problems that appear near the AH in FLRW cosmologies where
the LS develops the maximum area. The metric has its standard form
\begin{equation}\label{2-1}
ds^{2}=-dt^{2}+R^{2}(t)\Big( {dr^{2} \over 1 - \kappa r^{2} }
+r^{2}d\Omega_{2}^{2}\Big);
\end {equation}
the coordinates are comoving with the cosmic fluid, $t$ is the cosmic time and
$d\Omega_{2}^{2}$ is the metric of the two dimensional unit sphere. 
$\kappa=\pm 1,0$ is the spatial curvature and the velocity of light is
$c=1$. By the isotropy property the origin $r=0$ is a generic point and all
directions are equivalent; we fix the polar angles by fixing a radial
direction. We can use conformal coordinates 
\begin{equation}\label{2-2}
d\eta={dt \over R(t)}, \quad d\chi={dr \over (1-\kappa r^{2})^{1 \over 2}} ,
\end{equation}
and in this coordinates the metric is conformal to the flat metric,
\begin{equation}\label{2-3}
ds^{2}=R^{2}(\eta)(-d\eta^{2}+d \chi^{2}).
\end{equation}
When we analyze the causal structure in the unphysical flat metric, we do
not appreciate nothing particular at the location of the AH; however, if
we use proper coordinates\cite{ellis} to locate the events, a maximum of the
proper distance appears for the flat case (see Figure \ref{coord}). The
proper distance to the origin $D$ for an event at a given time, is relate to
its radial coordinate by
\begin{equation}\label{2-4}
dD^{2}=R^{2}(t) {dr^{2} \over 1-\kappa r^{2}},
\end{equation}
and the relevant part of the metric adopts the form
\begin{equation}\label{2-5}
ds^{2}=-dt^{2}+(dD-HDdt)^{2},
\end{equation}
where $H=H(t)$ is the Hubble constant. 
The null geodesics, that would constitute
the LS where we must measure the entropy, are given, in the conformal case
by $\eta=\pm \chi$; if we use proper coordinates, $ds^{2}=0$ translates into
\begin{equation}\label{2-6}
\dot D =HD \pm 1;
\end{equation}
$\dot D$ is the derivative of $D$ with respect to cosmic time.
The two signs correspond to the outgoing/ingoing nature of the photon with
respect to the origin. The AH appears for the ingoing light ray; the photon
near the Big Bang, although it is directed toward the origin (its radial
coordinate $r$ is decreasing), due to the expansion of the Universe begins 
to recede in proper coordinates; as the expansion evolves the proper
distance of the photon attains a maximum and subsequently begins to diminishes
until the origin is reached\footnote{It is possible that the photon 
never reaches the origin; this is a 
signal of the presence of event horizons.}.
The AH is locate where the area spanned by the ingoing photons is maximum. If
the universe is flat maximum area implies maximum proper distance $D$, so 
that $\dot D=0$, and substituting in (\ref{2-6}) the proper distance to 
the AH is given by $D_{AH}=1/H$.
In Figure \ref{coord} the past lightcone of a fiducial observer placed
at the origin is drawn for the two coordinates mentioned above, the conformal
$(\eta,\chi)$, and the proper $(t,D)$. The part of this lightcones beginning
in the apparent horizon and directed in both cosmic time directions conform
two LS where we study the validity of the GCEB.

\subsection{Violation of the GCEB near the AH}
The GCEB establish that the entropy traversing a truncated LS is bounded
by the difference of the areas of the two limiting spacelike surfaces.
In our analysis one of the boundaries will be the spacelike surface
spanned by the AH. We extend our LS up to another isotropic spacelike surface.
We assume a density of entropy $s(t)$ that depends only on cosmic time $t$,
and use proper coordinates $(D(t),t)$. For simplicity we develop the flat case
$\kappa=0$, but the result is easily generalized. We use units with 
$G=c=\hbar=k_{B}=1$.

The formulation of the GCEB is
\begin{equation}\label{2-7}
S(t)\leq{1 \over 4} \Delta A=  {1 \over 4}(A(t_{AH})-A(t)),
\end{equation}
where $A(t)$ is the area of the the isotropic surface bounding the LS, 
$t_{AH}$ is the cosmic time that locates the AH and $S(t)$ is the amount of
entropy passing the truncated LS. In terms of the proper distance $D(t)$ 
we have
\begin{equation}\label{2-8}
S(t)= \int_{t_{AH}}^{t} dt'  4 \pi D^{2}(t') s(t'),
\end{equation}
and
\begin{equation}\label{2-9}
{1 \over 4} \Delta A= \pi ( D^{2}(t_{AH})-D^{2}(t)).
\end{equation}
Taking $t$ near $t_{AH}$ and Taylor expanding (\ref{2-8}) and (\ref{2-9})
near the AH gives
\begin{equation}\label{2-10}
S\simeq 4 \pi D^{2}(t_{AH}) s(t_{AH}) \Delta t,
\end{equation}
and
\begin{equation}\label{2-11}
{1 \over 4} \Delta A \simeq - \pi D(t_{AH}) {\ddot D}(t_{AH}) \Delta t^{2},
\end{equation}
where $\Delta t= |t_{AH}-t|$.
It is then clear that for small enough values of $\Delta t$, 
${1 \over 4} \Delta A < S$ and the GCEB is violated.

\section{Restricting $\Delta t$ to satisfy the GCEB}
In order to respect the GCEB we must limit the separation in cosmic time
between both spacelike surfaces, so $\Delta t$ would be greater than
a minimum value $\Delta t_{m}$ that is obtained saturating the bound,
$S(\Delta t_{m})= \Delta A(\Delta t_{m})/4$,
\begin{equation}\label{2-12}
\Delta t_{m} =-4{D(t_{AH}) \over {\ddot D(t_{AH})}} s(t_{AH}).
\end{equation}
The equation of the null geodesic (\ref{2-6}) is now derived with respect 
to the cosmic time 
\begin{equation}\label{2-13}
{\ddot D} ={\dot H} D+ {\dot D} H, 
\end{equation}
and on the AH, for a flat universe, the maximum area implies maximum
distance ${\dot D}=0$; substituting in (\ref{2-13}) we obtain that
${\dot H}= {\ddot D}/ D$; then, using (\ref{2-12}), the GCEB is satisfied if 
\begin{equation}\label{2-14}
\Delta t \geq \Delta t_{m} = -4 {s(t_{AH}) \over { \dot H} (t_{AH}) }.
\end{equation}
For $\kappa \neq 0$ the previous expression generalizes to
\begin{equation}\label{2-15}
\Delta t \geq  -4 {s(t_{AH}) \over 
{ \dot H} (t_{AH})- {\kappa \over R^{2}(t_{AH}) }}.
\end{equation}

Our analysis has been purely kinematic, now we use the dynamics. The 
Friedman equations allows us to write
\begin{equation}\label{2-16}
{\dot H} -{\kappa \over R^{2}}= -4 \pi (\rho +p),
\end{equation}
where $\rho$ and $p$ are the density and pressure of the fluid that 
governs the cosmic evolution. The previous expression is valid also for
non zero cosmological constant. If the equation of state of the fluid is
$p= \omega \rho $, the restriction on the cosmic time can be put as
\begin{equation}\label{2-17}
\Delta t \geq {1 \over \pi(1+\omega)} {s \over  \rho},
\end{equation}
where the variables are evaluated on the AH. If the expansion is adiabatic
\begin{equation}\label{2-18}
s={ \rho+p \over T},
\end{equation}
where $T$ is the temperature of the fluid. Then (\ref{2-17}) adopts the form
\begin{equation}\label{2-19}
\Delta t \geq {1 \over \pi T_{AH}},
\end{equation}
the temperature being evaluated on the AH. 
This result was obtained by Bousso\cite{bousso3} for the 
particular example of Guedens\cite{guedens} (a 
closed radiation dominated universe). We see that (\ref{2-19}) is a general 
result, only requiring an adiabatic expansion.
\section{Interpreting the results}
We have obtained the  minimum value for the cosmic time 
that we must wait to satisfy
the GCEB, that is $\Delta t_m \sim s/\rho$. It is important to know 
the behavior of this value as the universe evolves; so we
study the quotient between
$\Delta t_m$ and a cosmic time scale. We know that if the expansion
is adiabatic and $\omega$ constant, 
$s \sim R^{-3}$ and $\rho \sim R^{-3(1+\omega)}$. On the other 
hand for a flat universe $R \sim t^{2 \over 3(1+\omega)}$. Taking $t_{AH}$ as
the scale we have
\begin{equation}\label{2-20}
{ \Delta t_m \over t_{AH}} \sim t_{AH}^{ \omega -1 \over \omega +1};
\end{equation}
it is clear that the previous quotient decreases with cosmic time if
$\omega <\omega_{c}=1$, the Fischler-Susskind limit\cite{fischler}.

Let us now focus on the expression (\ref{2-19}). If $T_{AH}$ is the temperature
of the cosmic fluid, its inverse will be a measure of the typical 
wavelength of the quanta that carries the entropy,
$\Delta t \geq 1/ \pi T_{AH} \equiv \lambda_{M}$. So, when we count the
entropy that traverses the LS, it is natural to consider only those modes 
whose wavelength is smaller than the size of the LS; that is, 
$\lambda < \lambda_{M}$ and we must cut the modes using the previous 
infrared cutoff \cite{bousso3}. If $t_{0}$ is the cosmic time where the 
LS, beginning in the AH, is truncated, when counting the amount of entropy 
passing by such LS we must use a thermal distribution with an IR cutoff 
$\lambda \leq \lambda_{M} =t_0 - t_{AH}$\footnote{ We consider the future 
directed LS, the past directed one admits a similar analysis and we give the 
numerical result  for both LS.};
the density of entropy at each time (or temperature) is obtained integrating
the distribution with the appropriate limits,
\begin{equation}\label{2-21}
s(T,\lambda_M)= {2 g T^3 \over 3 \pi^2 }
\int_{\lambda_{M}^{-1}}^{\infty} 
{ dx \, x^3 \over e^x -1};
\end{equation}
we suppose that the quanta is bosonic and that the degeneracy is $g$. Now 
we integrate the previous density between two temperatures (or times)
\begin{equation}\label{2-22}
S(t_f, \lambda_M)=
\int_{T_{AH}}^{T_{0}} dt(T) 4 \pi D^2 s(T,\lambda_M),
\end{equation}
where for the dependence of $D$ with $t$ we must integrate (\ref{2-6})
using the explicit function for the scale factor\cite{nos}.
The functional dependence of cosmic time with the temperature depends on 
the nature of the fluid that fills the universe; for radiation, $t \sim T^2 $
and we can compute the previous integral; it only remains to compare with 
one quarter of the decrease of the area. 
\begin{figure}[t]
\epsfxsize=20pc 
\epsfbox{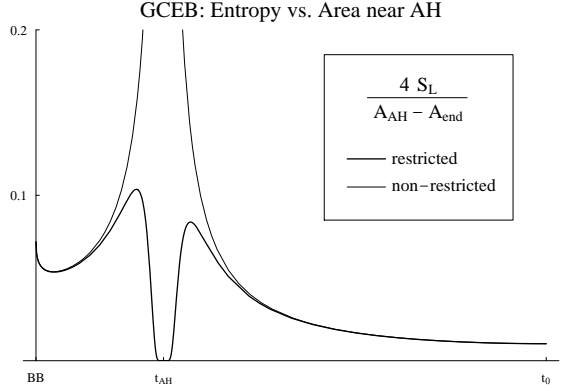} 
\caption{Validity and violation of the GCEB with and without an IR cutoff
respectively, for a radiation dominated universe.
\label{guedens}}
\end{figure}
In the Figure \ref{guedens} we plot this relation and we see that the 
introduction of the physical IR cutoff restores the GCEB for all truncated LS
on this sort of cosmological scenarios

The introduction of a cutoff to satisfy the GCEB imply a non-additivity
of the entropy. Consider a truncated LS $L$ made by two
adjacent LS's so that $L=L_1+L_2$; the entropy on $L_i$ accounts for 
the modes with $\lambda \leq \lambda_i$, $i=1,2$; for the entropy on $L$ 
we must consider the modes with $\lambda \leq \lambda_1+\lambda_2$.
It is clear that we lose the additivity (extensivity) of the entropy and 
\begin{equation}\label{2-23}
S(L=L_1+L_2) \geq S(L_1)+S(L_2).
\end{equation}
Wether this is a mathematical artifact or can be related with a fundamental non
additivity of the entropy in string/M-theory is an open question.

To finish the discussion of our results, consider 
the expression (\ref{2-17}); near the AH a constant density is a good 
approximation and we have
\begin {equation}\label{2-24}
\Delta t \geq {S \over M},
\end{equation}
where $S$ and $M$ are the total entropy and mass traversing the small LS. 
There is an adjusting mechanism; if more entropy tries to pass the LS, 
for a given $\Delta t$, more energy is used to carry the entropy and 
consequently the LS curves more and the bound is satisfied.
In the extreme situation, when only one bit of mass $M_{1}$
passes through the LS, $S \sim 1$ 
and (\ref{2-24}) has the form of a cosmological uncertainty relation, namely
\begin{equation}\label{2-25}
\Delta t \, M_{1} \geq 1,
\end{equation}
suggesting a deep connection between quantum mechanics, general covariance and
the GCEB, in this cosmological setup.

\section*{Acknowledgments} We acknowledge discussions with R. Bousso, 
R. Guedens and S.F. Ross.
We thanks also L. J. Boya for reading the manuscript.
This work was supported by MCYT (Spain) grant FPA2000-1252.

\end{document}